\documentclass[10pt, conference]{IEEEtran}
\IEEEoverridecommandlockouts
\usepackage{cite}
\usepackage{textcomp}
\usepackage[a4paper]{geometry}
\usepackage{marvosym}
\usepackage{CJKutf8}
\usepackage{changes}

\usepackage{gensymb}
\usepackage{bm}

\usepackage{amsmath,amssymb}
\usepackage{amsthm}

\usepackage{balance}

\usepackage{graphicx}
\graphicspath{{figures/}}
\usepackage[export]{adjustbox}
\usepackage{flushend}

\usepackage{subfigure}

\usepackage{listings}
\usepackage{color}
\usepackage{soul}
\usepackage{tikz}
\newcommand*{\circled}[1]{\lower.7ex\hbox{\tikz\draw (0pt, 0pt)%
    circle (.5em) node {\makebox[1em][c]{\small #1}};}}

\usepackage{algorithm}
\usepackage{algpseudocode}

\theoremstyle{definition}
\newtheorem{definition}{Definition}

\newtheorem*{proof}{Proof}
 
\theoremstyle{remark}

\definecolor{mygreen}{rgb}{0,0.6,0}
\definecolor{mygray}{rgb}{0.5,0.5,0.5}
\definecolor{mymauve}{rgb}{0.58,0,0.82}

\usepackage{algpseudocode}
\usepackage{algorithm}

\usepackage{booktabs} 
\usepackage{multirow}
\usepackage{array}
\newcolumntype{I}{!{\vrule width 1.2pt}}
\newlength\savedwidth

\newlength\savewidth

\usepackage{makecell}

\usepackage{etoolbox}
\makeatletter
\patchcmd{\@makecaption}
  {\scshape}
  {}
  {}
  {}
\makeatletter
\patchcmd{\@makecaption}
  {\\}
  {.\ }
  {}
  {}
\makeatother

\usepackage{geometry}
\newcommand{\defref}[1]{\mbox{Definition~\ref{#1}}}

\newcommand{\figref}[1]{\mbox{Fig.~\ref{#1}}}

\renewcommand{\eqref}[1]{\mbox{Equation~(\ref{#1})}} 

\usepackage[capitalise,nosort]{cleveref}      
\begin{document}
\begin{CJK*}{UTF8}{gbsn}

\title{Enhancing ASIC Technology Mapping via Parallel Supergate Computing}

\iftrue
\author{
\IEEEauthorblockN{
Ye Cai$^{1,*}$,
Zonglin Yang$^{1,*}$,
Liwei Ni\textsuperscript{2,3,\Letter},
Biwei Xie$^{3,2}$,
Xingquan Li\textsuperscript{2}
}\\

\IEEEauthorblockA{
$^1$College of Computer Science and Software Engineering, Shenzhen University, Shenzhen, China,\\
$^2$Peng Cheng Laboratory, Shenzhen, China,\\
$^3$State Key Lab of Processors, Institute of Computing Technology, Chinese Academy of Sciences, Beijing, China,\\
}
\IEEEauthorblockA{
Email: 
ycai@szu.edu.cn,~
2100271085@email.szu.edu,~
\textsuperscript{\Letter}nlwmode@gmail.com,~
xiebiwei@ict.ac.cn,~
lixq01@pcl.ac.cn
}
\thanks{*Equivalent contribution author.}
}

\fi

\maketitle

\begin{abstract}
With the development of large-scale integrated circuits, electronic design automation~(EDA) tools are increasingly emphasizing efficiency, with parallel algorithms becoming a trend. The optimization of delay reduction is a crucial factor for ASIC technology mapping, and supergate technology proves to be an effective method for achieving this in EDA tools flow. However, we have observed that increasing the number of generated supergates can reduce delay, but this comes at the cost of an exponential increase in computation time. In this paper, we propose a parallel supergate computing method that addresses the tradeoff between time-consuming and delay optimization.
The proposed method utilizes the input-constrained supergate pattern to parallelly generate the supergate candidates, and then filter the valid supergates as the results. Experiment results show the efficiency of the proposed method, for example, it can attain the improvement of 4$\times$ speedup in computation time and 10.1 in delay reduction with 32 threads.
\end{abstract}


\section{Introduction}
\label{sec:intro}

With the development of large-scale integrated circuits, EDA tools are required to employ more efficient algorithms that can run on faster processors. 
The trend towards parallel computing has emerged as a solution, leveraging the capabilities of widely available multi-core platforms to enhance the efficiency of EDA tools. 
As a result, numerous parallel algorithms have been developed for various stages of the EDA process~\cite{para2, para3, para4, lasted-GPU-unlock, liu2023rethinking}.
Technology mapping is an essential step in the logic synthesis of EDA flow for digital systems, such as ASICs and FPGAs. 
It involves converting a technology-independent logical network into a functionally equivalent circuit implemented using primitive gates selected from a technology library. 
The primary objective is to minimize some criterion, such as area, power, and delay~\cite{chatterjee2007algorithms}.
As the problems of energy consumption and heat dissipation of digital circuits become increasingly obvious, delay becomes critical\cite{para1}.
The increasing complexity of SoC design and the market pressure to reduce time to market, make the systems designer's task harder especially for ASIC design\cite{ASIC-time}.
However, technology mapping becomes time-consuming due to the booming scale and complexity of IC designs\cite{ASIC-time2}. 

The process of ASIC technology mapping involves the translation of the logic network into a predefined set of standard cells. 
In the developments, there has been a shift in ASIC technology mapping from tree-based mapping \cite{tree-mapping} to cut-based mapping \cite{priority-cut}. 
Cut-based methods offer enhanced exploration of diverse logic structures and optimization, particularly with scalability to large-scale designs.
The limited standard cells in the standard cell library can not meet all the cut functions, which makes some possibilities lost in the mapping process.
The emergence of supergate technology enables the standard cell library to display more structural composite gates to achieve more functions to meet the needs of mapping and to achieve a lower delay\cite{supergate-2005}.

Through the preliminary experiment, we have observed that the change in the generated supergates size will affect the quality of the ASIC mapping results.
Especially, increasing the number of generated supergates can reduce delay, but this comes at the cost of an exponential increase in computation time.
Also, the supergate computing step occupies most of the time of ASIC technology mapping in practice.
Parallelism is an effective approach to reducing the cost of time.
However, the challenge is how to make the supergate computing parallel and how to solve the data conflict.

To address this challenge above, in this paper, we propose a parallel supergate computing method that addresses the tradeoff between time-consuming and delay optimization.
The proposed method utilizes the input-constrained supergate pattern to parallelly generate the supergate candidates, and then filter the valid supergates as the results.
By parallelising the most time-consuming parts, we greatly accelerate the speed of supergate generation, and by limiting the number of attempts the cut corresponding supergates in the mapping phase, we can balance the time and quality.
Experiment results show the efficiency of the proposed method, for example, it can attain the improvement of 4$\times$ speedup in computation time and 10.1\% in delay reduction with 32 threads.
The contributions can be summarized in the following three folds:
\begin{itemize}
    \item We analyze the supergate computing procedure and figure out the conflict problem between the computation time and the generated supergate size;
    \item We propose a parallel supergate computing method that utilizes the input-constrained supergate pattern to parallelly generate the supergate candidates;
    \item The sufficient experiments have been performed to show the efficiency of the proposed method with runtime and delay reduction.
\end{itemize}

The rest of this paper is structured as follows. \cref{sec:pre} provides the basic notations and concepts of ASIC technology mapping and supergates.
In \cref{sec:motivation}, we give the motivation and problem formulation.
\cref{sec:method} shows the details of the proposed parallel supergates computing.
The \cref{sec:exper} and present the experiments result and the conclusion.

\section{Preliminaries}
\label{sec:pre}
In this section, we provide the basic notations and the basic concepts of ASIC technology mapping and supergate computing.

\subsection{Basic Notations}
\paragraph{Boolean Function}
An $n$-variable Boolean function $f(x)$ takes the form $f: \{0, 1\}^{n} \rightarrow \{0, 1\}$, where $\{0, 1\}$ is the Boolean domain and $n$ is the number of Boolean variables $x$ of $f$.
In practice, the Boolean function $f$ is often represented by the truth table $T(f)$, a binary string with $2^{n}$ bits. 

\begin{figure}[t]
    \centering
    \includegraphics[width=0.45\textwidth]{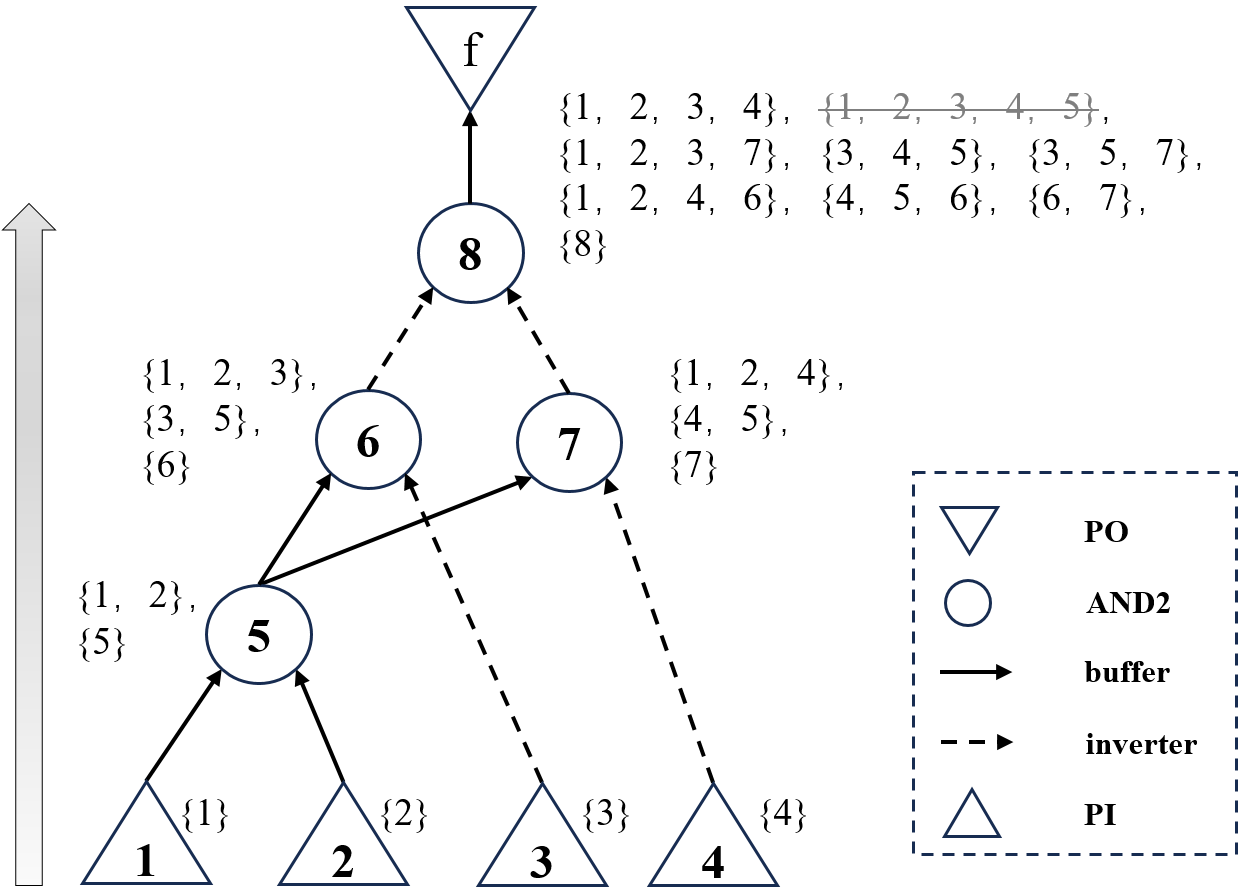}
    \caption{The illustration of the 4-feasible cut enumeration procedure on an AIG format Boolean network. Each node is assigned the trivial cut first, then the iteratively merging operation is performed following the bottom-up direction.}
    \label{fig:aig_with_cuts}
\end{figure}

\paragraph{And-Inverter Graph} 
An And-Inverter Graph~(AIG) is a directed acyclic graph~(DAG), which is the commonly used Boolean representation in logic synthesis.
An AIG can be formulated by $\mathcal{G}=(\mathcal{V}, \mathcal{E})$, where the node set $\mathcal{V}$ consists of the input nodes~(PIs), output nodes~(POs) and the internal AND2 nodes; 
And the edge set $\mathcal{E}$ consists of the buffers and inverters, where buffer refers to the signal and inverter refers to the complemental signal.
The \figref{fig:aig_with_cuts} illustrate the structure of an example of AIG.

\begin{figure}[t]
    \centering
    \includegraphics[width=1.0\linewidth]{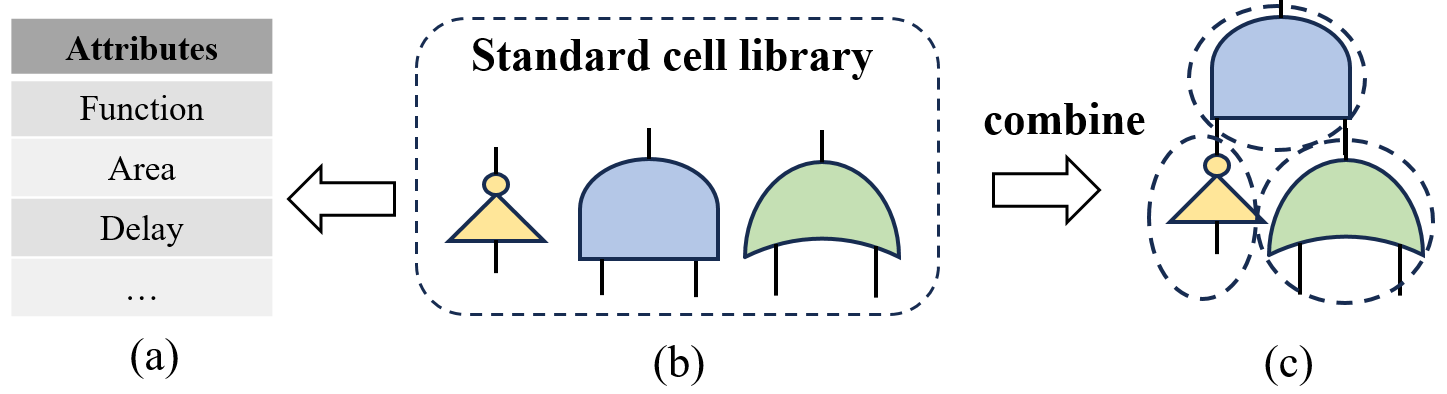}
    \caption{The illustration of the standard cell and supergate computing. The (a) and (b) show the attributes of the standard cells, while (b) and (c) show the example of supergate computing.}
    \label{fig:prelim-supergate}
\end{figure}

\paragraph{Standard Cell}
Standard cells are fundamental building blocks used in integrated circuit (IC) design.
The main purpose of standard cells is to provide a library of reusable components with fixed functions and specifications. 
Each standard cell has a specific function and physical attributes. 
The specific function is defined by the Boolean expressions or the truth table; 
The physical attributes are defined by the foundry according to their fabrication technique, it consists of the area, delay, capacitance, resistance, and \textit{etc}.
The \figref{fig:prelim-supergate}(a) and (b) illustrate the standard cell.

\paragraph{Supergate} 
A supergate is a multi-input and single-output logic gate that consists of standard cells.
In the cut-based ASIC technology mapping process, the limited size of standard cells will limit the cut ranking and selection, which can affect the quality of the results~(QoR).
And the supergate technology can improve the number of cut matching by merging the standard cells to generate the new logic gate, which can enhance the cut ranking and selection to improve QoR.
And \figref{fig:prelim-supergate}(b) and (c) show an example of the supergate computing scheme. 

\subsection{Basic Concepts}

\paragraph{ $k$-feasible Cut} 
The cut $C$ of a node $v$ is a set of nodes such that every path from primary inputs (PIs) to $v$ must pass through at least one node of cut $C$.
A \textit{trivial cut} of node $v$ is the node itself. 
A \textit{non-trivial cut} includes all the nodes that are found on the paths from the root to the leaves.
A cut $C$ is $k$-feasible only if the number of nodes in it does not exceed $k$.

Let $\Phi(n)$ be the set of $k$-feasible cuts for node $n$, the \textit{merge} operation $\diamond$ of two cut sets for any given node $n_i$ and $n_j$ can be defined as the following: 

\begin{equation}
\label{eq:cut_merge}
    \begin{aligned}
        \Phi(n_i) \diamond \Phi(n_j) = \{u \cup v | u \in \Phi(n_i), v \in \Phi(n_j)\}.
    \end{aligned}
\end{equation}

And the \textit{cut enumeration} operation is based on the \textit{merge} operation $\diamond$ as shown in \eqref{eq:cut_merge}, and it can be formulated by the following:
\begin{equation}
\label{eq:cut_enumeraiton}
    \begin{aligned}
        \Phi(n) = \left\{
        \begin{aligned}
        \{\{n\}\} & :n\in PIs \\
        \{\{n\}\} \cup \Phi(n_1) \diamond \Phi(n_2)  & :otherwise
        \end{aligned}
        \right \},
    \end{aligned}
\end{equation}
where $n_1$ and $n_2$ are the two fanin nodes of node $n$, and the computation procedure follows the PIs to POs direction.

The \figref{fig:aig_with_cuts} also illustrates the 4-feasible cut enumeration procedure which is performed according to \eqref{eq:cut_enumeraiton}.
Constrained by the 4-feasible cut, the cut \{1, 2, 3, 4, 5\} of node 8 will be filtered out due to its size exceeding 4.

\paragraph{Boolean Matching.}
The Boolean matching method in ASIC technology mapping is to bind the cut with the computed standard cell or supergates.
An NPN~(inputs \textbf{N}egation, inputs \textbf{P}ermuation, output \textbf{N}egation) transformation is the commonly used method to expand the matching cut by allowing the inputs negation, inputs permutation, and output negation.
Through this technology, a cut $C$ is allowed to match several supergates to expand the space of cut selection.


\begin{figure}[t]
    \centering
    \includegraphics[width=1.0\linewidth]{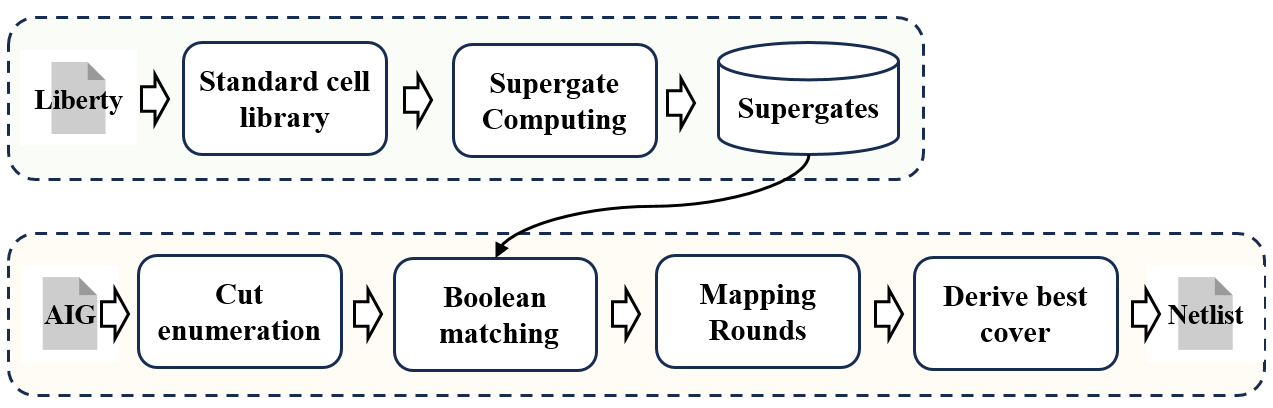}
    \caption{The illustration of ASIC technology mapping flow with the support of the supergate computing flow targeting the AIG circuits.}
    \label{fig:techmap-flow}
\end{figure}

\section{Motivation and Problem Formulation}
\label{sec:motivation}

\begin{figure*}[t]
    \centering
    \includegraphics[width=0.95\linewidth]{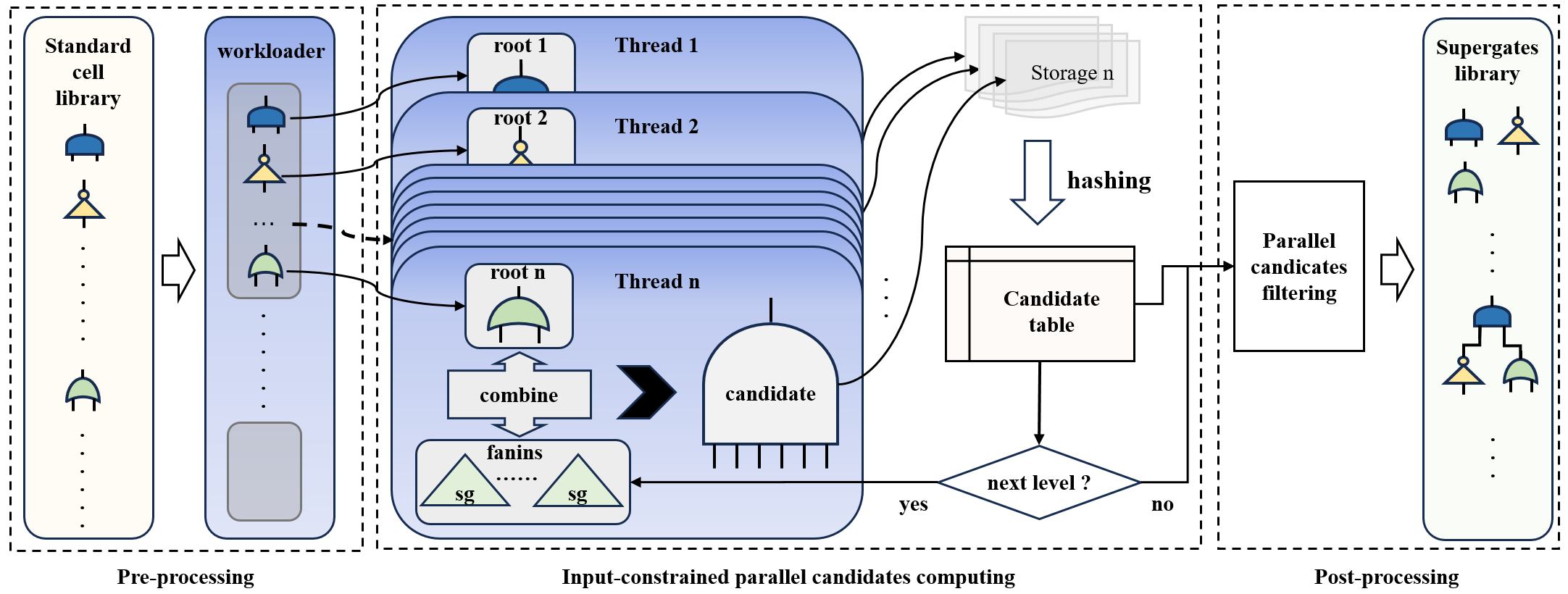}
    \caption{The framework of the proposed parallel supergate computing.}
    \label{fig:method-flow}
\end{figure*}

\begin{figure}[t]
\centering
    \subfigure[Delay reduction by supergate increase.]{ 
        \begin{minipage}[t]{0.45\textwidth}
        \centering
        \includegraphics[width=\textwidth]{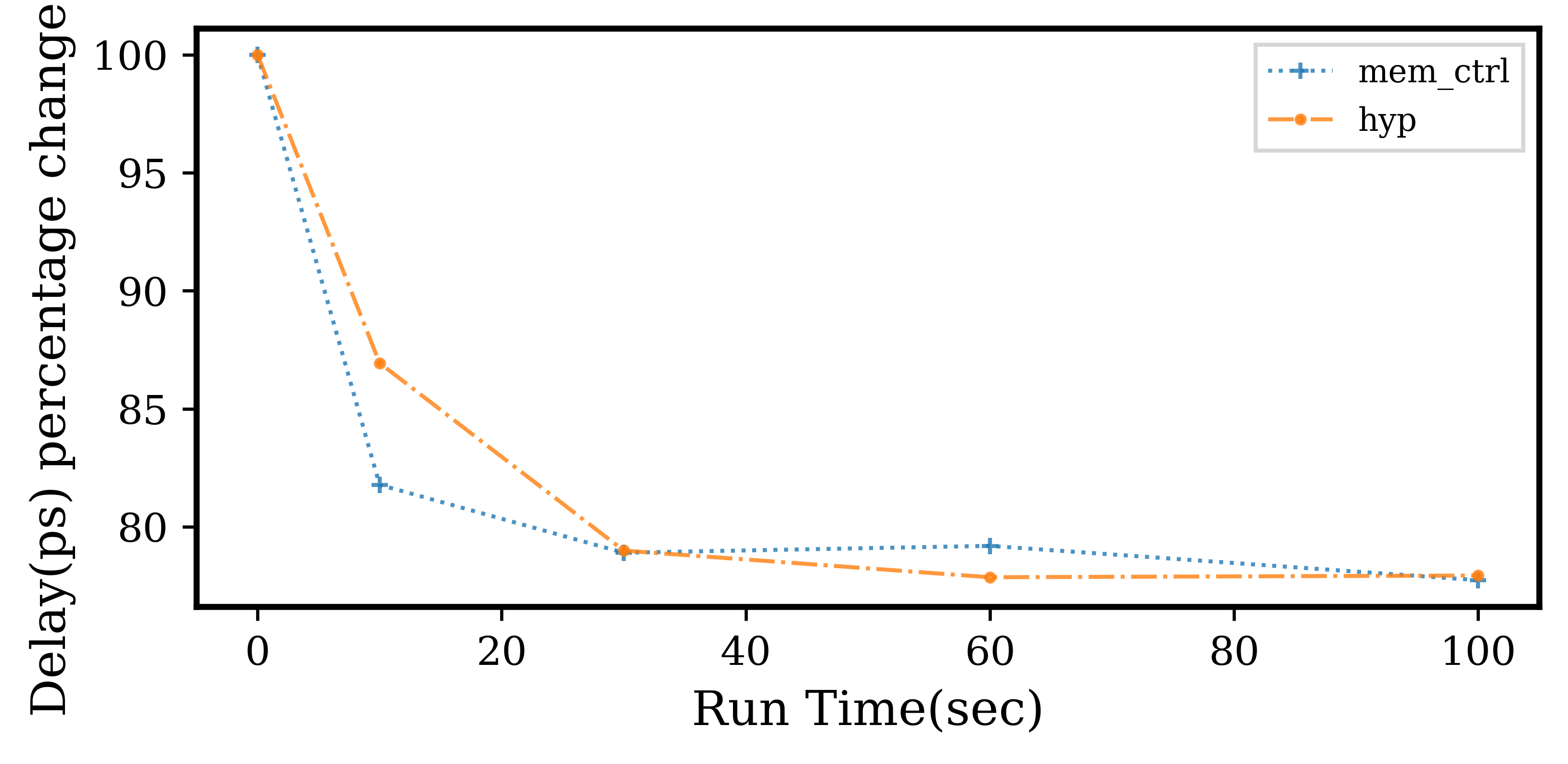}
        \end{minipage}
        \label{fig:motivation_exper:a}
    }
    \hfill
    \subfigure[Supergates Generation by sky130 liberty file.]{ 
        \begin{minipage}[t]{0.45\textwidth}
        \centering
        \includegraphics[width=\textwidth]{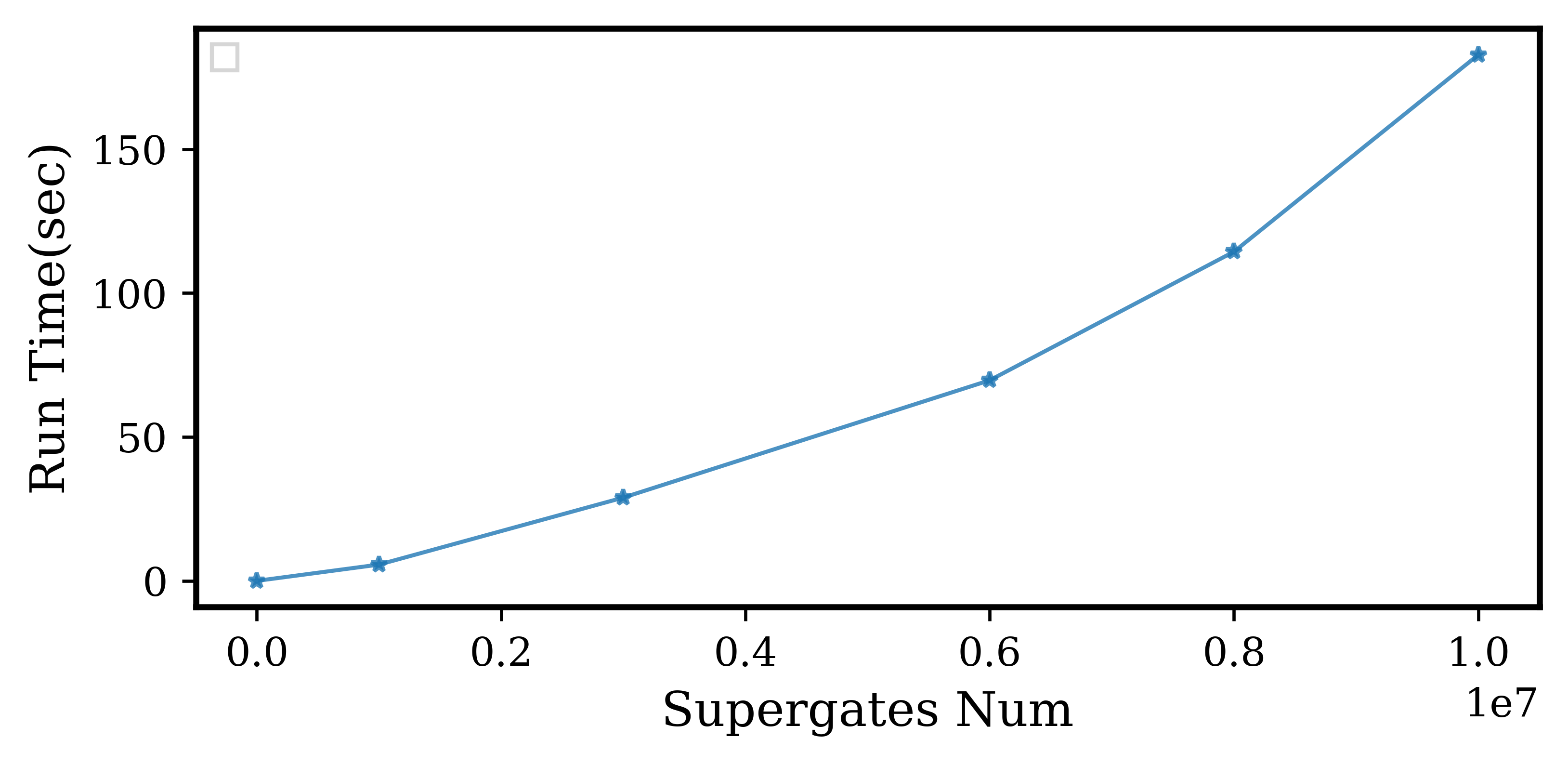}
        \end{minipage}
        \label{fig:motivation_exper:b}
    }
    \caption{The illustration of the experiment results.}
    \label{fig:motivation_exper}
\end{figure}

In this section, we will outline the basic flow for ASIC technology mapping and supergate computing.
Then, the motivation and problem formulation will be given.

\subsection{ASIC Technology Mapping Flow}

The \figref{fig:techmap-flow} illustrates the ASIC technology mapping flow with the support of supergate computing targeting the AIG circuit.
As for the supergate computing flow, the input liberty file will be parsed into the set of standard cells, then the supergate computing step will generate the supergates data from the standard cells. 
As for the ASIC technology mapping flow, the AIG file and the generated supergates data as the input files.
The first step is to perform the cut enumeration to generate the local slices of AIG;
The following Boolean matching technology binds the cuts of each node with the supergates, which can assign the Boolean function and physical attributes of the cut to carry on the following cut ranking and cut selection.
Then the multi-round mapping will be performed by different criteria.
Finally, the gate-level netlist will be derived by covering the AIG with the best cut of each node from POs to PIs.

\subsection{Motivation Experiment}
The following two experiments are performed to show the supergate's effect on ASIC technology mapping. 
 By changing the runtime of supergate computing to limit the number of generated supergates, the \figref{fig:motivation_exper:a} illustrates the effect on delay reduction for the given AIG, and the cases are from EPFL dataset~\cite{benchmark}.

By limiting the number of generated supergates, the runtime line of different numbers of supergates is shown in \figref{fig:motivation_exper:b}, and it is clear that the runtime grows faster as the supergates increase. 

\subsection{Problem Formulation}
The above motivation experiments show the conflict between the computation time and delay reduction of the supergate computing.
So the problem of this paper is to make the supergate computing parallel.

\section{Proposed Methodology}
\label{sec:method}


\subsection{Overview}


The \figref{fig:method-flow} shows the framework of the proposed method, and it can be roughly divided into the following three steps:
\paragraph{Pre-processing}
This step is to initialize the input-constrained supergate storage model and the hashing tables, the workloader will also be initialized.
\paragraph{Input-constrained parallel candidates computing}
This step is to parallelly combine the standard cell into the candidate supergates. Due to the exponential space, the level~(the depth by combining different gates) will be the terminal condition to end the procedure.
\paragraph{Post-processing}
This step is to parallelly filter out the candidate supergates, only the valid candidates will be put into the final supergates library.







\subsection{Parallel Supergate Computing}
\subsubsection{Pre-processing}

Several pre-processing steps need to be performed before parallel computing.

\begin{figure}[t]
    \centering
    \includegraphics[width=0.46\textwidth]{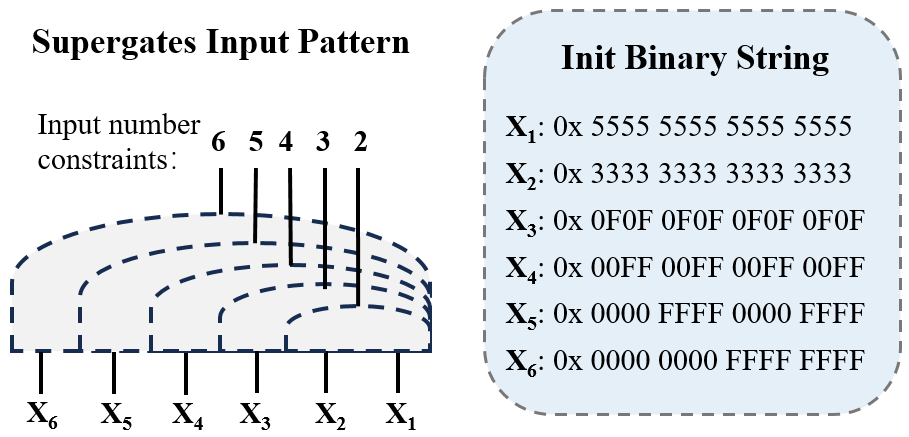}
    \caption{The illustrations of the input-constrained supergates pattern and the binary string initialization. The truth table of each supergate is simulated by the initial binary string of the corresponding Boolean variables. In practice, the length $len$ of the truth table is constrained by the input number $n$ with $len = 2^n$.}
    \label{fig:supergate-pattern}
\end{figure}

The first step is to construct the initial supergate set by the input-constrained supergate pattern as shown in \figref{fig:supergate-pattern}.
The supergates in the initial set do not contain the actual standard cells. and their number is equal to the largest input number~(6 in practice) of the gate supported by the library.
The truth table also corresponds to the value of the binary variable.
As shown in \figref{fig:supergate-pattern}, there are 6 binary variables, $x_1$ to $x_6$. and each Boolean variable will be assigned the corresponding binary string.
The truth table of combined supergates according to the pattern can be simulated by the corresponding Boolean variable's binary string as the input signal for the internal standard cells. 








Next, we will introduce the storage mode of data and explain its purpose. 
We use queues to store candidates and a hash table to store supergates. 
To support multi-threaded storage of candidates, we have several storage queues, each with a queue for threads. 
Mutex locks are used to avoid conflicts. We map the truth table to a smaller range of keys and initialize the corresponding mutexes.
Since the candidates generated by multiple threads may have the same truth table in a multithreaded environment, it is necessary to use mutex locks to avoid conflicts.
The range of truth tables is very large, for example, the range of the truth table of 6 variables is $2^{2^6} = 2^{64}$.
To save time and space, map the truth table to a smaller range of keys, and then initialise the corresponding number of mutexes.
The gates that can be used as roots are selected from the standard cells, the root gates of the same function in the standard cells will only leave the one that has the lowest delay.

Finally, the standard cell that can be the root of supergate will be loaded into the workloader by thread limits.

\subsubsection{Input-constrained parallel candidates computing}
The following is the main body of the Parallel Compute Candidates algorithm for computing the supergates of each round.
As shown in \figref{fig:method-flow}, the generation of each new round of supergate candidates depends on the supergate candidates generated in the previous round. 

\begin{algorithm}[t]
\small
\caption{Parallel Candidates Computing}
\label{algorithm:method-parallel-compute}
\begin{algorithmic}[1]
\Require candidates set $C$, hash table $H$, root gates $roots$, inputs size $k$, prune limit $limit$, threads num $N$

\Statex /* collect fanin gates candidates in $C$, which are smallest delay in same function */ 

\State $F$ $\leftarrow$ collect\_fanin\_gates\_candidates($C$)

\ForAll{$r \in \text{parallel range}(1, |roots|, N)$}

\If{$r$.inputs\_num $>$ $limit$ }:
   \State \textbf{continue}
\EndIf

\Statex /* collect fanin gates candidates for root $r$ */ 
\State $f$ $\leftarrow$ collect\_fanin\_gates($F$, $r$)

\Statex /* recursive enumerate fanins permutations */ 
\State $fa \gets \text{new array}[k]$
\State $idx \gets \text{0}$
\State recur\_enum\_perm($C$, $H$, $r$, $f$, $fa$, $idx$)

\EndFor

\end{algorithmic}
\end{algorithm}

The \cref {algorithm:method-parallel-compute} initially identifies the gate with the minimal delay from the supergate candidates of the same function in $ C $ as the candidate for the fanin gates (line 1).
All roots satisfying the input constraints are concurrently processed (line 2).
The procedure begins by gathering fanin gate candidates that are compatible with the current root $ r $, followed by employing a recursive algorithm to exhaustively enumerate all possible combinations among them (line 6).


\begin{algorithm}[t]
    \small
    \caption{Recursive Enumerate Combinations}
    \label{algorithm:method-recursive}
    \begin{algorithmic}[1]
        \Require candidates set $C$, hash table $H$, root gates $roots$, fanins gates $fanins$, fanins array $fa$, index in fanins array $idx$
        
        \If{$fn == r.inputs\_num$}:
            \State $a$, $d$, $f \leftarrow compute\_composed\_data(r, fa)$
            \If{is\_best($a$, $d$, $f$)}:
                \State $C, H \leftarrow create\_new\_candidate(r, fa, a, d, f)$
            \EndIf
        \Else
            \For{$f$ in $fa$}
                \If{!is\_repetitive()}:
                    \State $fa$[$idx$] $\leftarrow$ $f$
                    \State recur\_enum\_perm($C, H, r, fanins, fa, idx$)
                \EndIf
            \EndFor
        \EndIf
    \end{algorithmic}
\end{algorithm}

The \cref{algorithm:method-recursive} is a method based on backtracking and recursion, which is used to enumerate all combinations of a root gate and several fanins gates.
The $fanins$ is all available fanins gates.
The current fanin gates combination is stored in $fa$, and $idx$ is the position in the fanins array to be modified.

Before a new supergate candidate is created, need to calculate its area, delay and truth table, and then compare whether it is the smallest delay in the same function in the hash table.
Note that there may be multiple optimal supergate candidates for the same function, and all of them need to be preserved.
The successfully created supergate candidate is stored in the queue exclusive to each thread, and their pointers are added to the hash table.
Here, the operation of the hash table needs to obtain the corresponding mutex to avoid conflict.

\subsubsection{Post-processing}
This step parallelly filters out the candidate supergates, only the valid candidates will be put into the final supergates library.
The following two definitions are under the constraints of \cref{fig:supergate-pattern}:
\begin{definition}[continuous-input candidate]
\label{def:valid_supergate}
The $n$-input candidate supergate is a continuous-input candidate only if the Binary variables are continuous and starting from $X_1$ to $X_n$.
\end{definition}

\begin{definition}[discontinuous-input candidate]
\label{def:unvalid_supergate}
The $n$-input candidate supergate is a discontinuous-input candidate only if the Binary variables are not continuous, or it is not starting from $X_1$.
\end{definition}

The supergate requires that the truth table is valid after the computing, thus it can be used for the Boolean matching steps.
The parallel candidates computing step will generate the two defined types of candidates: the continuous-input and discontinuous-input candidates.
Each of them can help to compute the candidates with more inputs, and the definition of the continuous-input and discontinuous-input candidate is the criterion to filter the valid supergate.
As for the continuous-input candidate, it can compute the correct truth table with the continuous binary string starting from $X_1$;
However, as for the discontinuous-input candidate, the input binary string of the Boolean variables are not following the correct input signal, the computed truth table is also wrong.

And we also give a parallel method to filter the valid supergates.
Since the storage mode of supergate candidates is queue storage with several threads, the filtering steps can also be carried out in parallel.
However, in the end, it is necessary to add the pointer of final supergate candidates to the library, and the library is also a truth table indexed to the hash table of corresponding supergates, so it is also necessary to re-hash and mutex lock to avoid multi-threaded write conflicts.

Due to that the parallel candidates computing procedure can generate a lot of supergates, the number of indexes in each truth table will also be very large. 
To balance efficiency and quality, it is necessary to limit the number of overrides attempted by each cut in the depth-oriented mapping and area recovery stages. 
This can sacrifice a small part of the candidates in exchange for a huge speed increase.

\subsection{Complexity Analysis}
Let $SG(l)$ be the set of candidate supergates of level $l$, then the supergates computing procedure can be formulated as follows:
\begin{equation}
\label{eq:sg}
|SG(l)| = \left\{
    \begin{aligned}
        k,       &&  (l = 1); \\
         \sum_{rg\in RG} \sum_{i=1}^{n_{rg}}A_{|SG(l-1)|}^i, && (l > 1),
        \end{aligned}
    \right.
\end{equation}
If the level of supergate is 1, the number of supergates in the set is $k$, $k$ is the maximum number of inputs of the gates supported by the library.
If the level is greater than 1, the calculation process of this set corresponds to the above formula, where the $\sum$ refers to the combining operation, Let $rg\in RG$ is root gate, and $RG$ be the set of root gates, and $n_{rg}$ refers to the size of input gate to $rg$. 
$A_{|SG(l-1)|}^i$ is permutation notation. From \eqref{eq:sg}, we can see that the size of $SG(l)$ is very huge for each level $l$.





\subsection{Case Study}
\label{case-study}

\begin{figure}[t]
    \centering
    \includegraphics[width=0.95\linewidth]{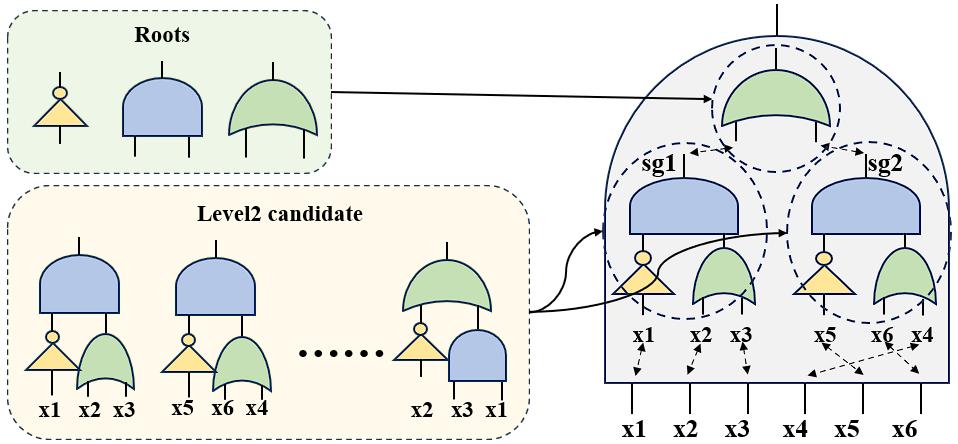}
    \caption{The illustration of the supergate computing with different levels, and different gates can be used as roots to generate candidates in parallel. The right of this figure shows a level-3 candidate generation with 6 inputs.}
    \label{fig:case}
\end{figure}

\cref{fig:case} shows a case study of the above parallel supergate computing procedure.
Under our method, each root gate corresponds to the execution of a thread. 
All threads share the candidates generated in the previous round, filter the available candidates as fanin gates, and then enumerate all possible patterns that can be composed with root gates.

In the process of combination, due to the enumeration method, based on the \defref{def:valid_supergate}, it is inevitable that invalid supergates will appear.
As shown in \figref{fig:case}, the two fan gates, $sg_1$ and $sg_2$, which constitute the 6 input candidates, have the same structure.
However because they have different elementary variables, their truth tables are also different. 
The expression of $sg_1$ is $f(sg_1) = (\neg x_1 \wedge (x_2 \vee x_3))$, and the truth table calculated by bit simulation is 0x 0E0E 0E0E 0E0E 0E0E. It is a valid supergate in itself. 
The expression of $sg_2$ is $f(sg_2) = (\neg x_5 \wedge (x_6 \vee x_4))$, and the calculated truth table is 0x 00FF 0000 FFFF 0000, which is not a supergate of correct function by itself.

But when $sg_1$ and $sg_2$ compose this new candidate, its overall element variables become $x_1$ to $x_6$, which is a continuous segment starting from $x_1$. 
Its truth table is $f(sg_1)\wedge f(sg_2) = $ 0x 0EFF 0E0E FFFF 0EFE, which is the correct truth table with the structure.

\section{Experimental Results}
\label{sec:exper}

\begin{table*}[t]
	\centering
	\caption{Our method compares with the results of ABC while not using supergate and using supergate respectively. The data in the table is the optimized percentage of area and delay, and the calculation method is imp=(ABC-ours)/ABC*100. The ($A$)($B$) pattern means the improvements and $A$ means the improvement compared with "ABC without sg", while $B$ means the improvement compared with "ABC with sg". }
	\resizebox{\linewidth}{!}{
		\label{table: QoR}
		\begin{tabular}{|c|rr|rr|rr|rr|rr|rr|}
			\hline
			\multirow{2}{*}{Benchmarks} & \multicolumn{2}{c|}{ABC without sg} & \multicolumn{2}{c|}{ABC with sg}   & \multicolumn{2}{c|}{4 threads } & \multicolumn{2}{c|}{8 threads} & \multicolumn{2}{c|}{16 threads}                      & \multicolumn{2}{c|}{32 threads}                                                                                                                                                                                                                                                                                                                                                \\ \cline{2-13}
			                            & \multicolumn{1}{r|}{Area}      & \multicolumn{1}{r|}{Delay}      & \multicolumn{1}{r|}{Area}       & \multicolumn{1}{r|}{Delay}     & \multicolumn{1}{r|}{Area imp\%}                      & \multicolumn{1}{r|}{Delay imp\%}                      & \multicolumn{1}{r|}{Area imp\%}                      & \multicolumn{1}{r|}{Delay imp\%}                      & \multicolumn{1}{r|}{Area imp\%}                      & \multicolumn{1}{r|}{Delay imp\%}                      & \multicolumn{1}{r|}{Area imp\%}                      & {Delay imp\%}                     \\ \hline
			priority                    & \multicolumn{1}{r|}{4433}      & \multicolumn{1}{r|}{16861.21}   & \multicolumn{1}{r|}{5215}       & \multicolumn{1}{r|}{10919.56}  & \multicolumn{1}{r|}{(-65.88)(-41.00)}                & \multicolumn{1}{r|}{(\textbf{36.48})(\textbf{1.91})}  & \multicolumn{1}{r|}{(-68.36)(-43.11)}                & \multicolumn{1}{r|}{(\textbf{35.02})(-0.34)}          & \multicolumn{1}{r|}{(-69.57)(-44.15)}                & \multicolumn{1}{r|}{(\textbf{34.13})(-1.71)}          & \multicolumn{1}{r|}{(-69.77)(-44.31)}                & (\textbf{36.48})(\textbf{1.91}  ) \\
			cavlc                       & \multicolumn{1}{r|}{2099.51}   & \multicolumn{1}{r|}{1546.01}    & \multicolumn{1}{r|}{2354.76}    & \multicolumn{1}{r|}{1360.86}   & \multicolumn{1}{r|}{(-13.83)(-1.49)}                 & \multicolumn{1}{r|}{(\textbf{17.08})(\textbf{5.80})}  & \multicolumn{1}{r|}{(-8.70)(\textbf{3.08})}                   & \multicolumn{1}{r|}{(\textbf{9.31})(-3.03)}           & \multicolumn{1}{r|}{(-10.13)(\textbf{1.81})}         & \multicolumn{1}{r|}{(\textbf{10.50})(-1.67)}          & \multicolumn{1}{r|}{(-9.06)(\textbf{2.76})}          & (\textbf{17.08})(\textbf{5.80}  ) \\
			arbiter                     & \multicolumn{1}{r|}{39769.39}  & \multicolumn{1}{r|}{7848.5}     & \multicolumn{1}{r|}{50667.34}   & \multicolumn{1}{r|}{7765.75}   & \multicolumn{1}{r|}{(-23.97)(\textbf{2.7})}                  & \multicolumn{1}{r|}{(\textbf{30.71})(\textbf{29.98})} & \multicolumn{1}{r|}{(-24.06)(\textbf{2.62})}         & \multicolumn{1}{r|}{(-15.38)(-16.61)}                 & \multicolumn{1}{r|}{(-23.96)(\textbf{2.70})}         & \multicolumn{1}{r|}{(-8.03)(-9.18)}                   & \multicolumn{1}{r|}{(-27.22)(\textbf{0.15})}         & (\textbf{30.71})(\textbf{29.98} ) \\
			i2c                         & \multicolumn{1}{r|}{4320.39}   & \multicolumn{1}{r|}{1600.66}    & \multicolumn{1}{r|}{4379.2}     & \multicolumn{1}{r|}{1784.14}   & \multicolumn{1}{r|}{(-6.55)(-5.11)}                  & \multicolumn{1}{r|}{(\textbf{14.72})(\textbf{23.49})} & \multicolumn{1}{r|}{(-6.02)(-4.60)}                  & \multicolumn{1}{r|}{(\textbf{14.13})(\textbf{22.96})} & \multicolumn{1}{r|}{(-5.53)(-4.11)}                  & \multicolumn{1}{r|}{(\textbf{14.88})(\textbf{23.63})} & \multicolumn{1}{r|}{(-6.43)(-5.00)}                  & (\textbf{14.72})(\textbf{23.49} ) \\
			voter                       & \multicolumn{1}{r|}{73127.63}  & \multicolumn{1}{r|}{6770.27}    & \multicolumn{1}{r|}{82821.93}   & \multicolumn{1}{r|}{7135.53}   & \multicolumn{1}{r|}{(\textbf{7.07})(\textbf{17.95})} & \multicolumn{1}{r|}{(\textbf{10.87})(\textbf{15.43})} & \multicolumn{1}{r|}{(\textbf{7.17})(\textbf{18.04})} & \multicolumn{1}{r|}{(\textbf{9.33})(\textbf{13.98})}  & \multicolumn{1}{r|}{(\textbf{6.93})(\textbf{17.82})} & \multicolumn{1}{r|}{(\textbf{9.75})(\textbf{14.37})}  & \multicolumn{1}{r|}{(\textbf{8.06})(\textbf{18.82})} & (\textbf{10.87})(\textbf{15.43} ) \\
			int2float                   & \multicolumn{1}{r|}{730.7}     & \multicolumn{1}{r|}{1041.1}     & \multicolumn{1}{r|}{765.73}     & \multicolumn{1}{r|}{985.7}     & \multicolumn{1}{r|}{(-8.22)(-3.27)}                  & \multicolumn{1}{r|}{(\textbf{15.09})(\textbf{10.32})} & \multicolumn{1}{r|}{(-7.02)(-2.12)}                  & \multicolumn{1}{r|}{(\textbf{11.00})(\textbf{5.99})}  & \multicolumn{1}{r|}{(-7.02)(-2.12)}                  & \multicolumn{1}{r|}{(\textbf{11.00})(\textbf{5.99})}  & \multicolumn{1}{r|}{(-7.70)(-2.78)}                  & (\textbf{15.09})(\textbf{10.32} ) \\
			ctrl                        & \multicolumn{1}{r|}{479.21}    & \multicolumn{1}{r|}{687.65}     & \multicolumn{1}{r|}{497.98}     & \multicolumn{1}{r|}{845.97}    & \multicolumn{1}{r|}{(-0.52)(\textbf{3.27})}          & \multicolumn{1}{r|}{(\textbf{3.49})(\textbf{21.55})}  & \multicolumn{1}{r|}{(-0.52)(\textbf{3.27})}          & \multicolumn{1}{r|}{(\textbf{3.48})(\textbf{21.54})}  & \multicolumn{1}{r|}{(-0.52)(\textbf{3.27})}          & \multicolumn{1}{r|}{(\textbf{6.14})(\textbf{23.70})}  & \multicolumn{1}{r|}{(-0.78)(\textbf{3.02})}          & (\textbf{3.49})(\textbf{21.55 })  \\
			dec                         & \multicolumn{1}{r|}{1186.14}   & \multicolumn{1}{r|}{635.01}     & \multicolumn{1}{r|}{1186.14}    & \multicolumn{1}{r|}{635.61}    & \multicolumn{1}{r|}{(0.00)(0.00)}                    & \multicolumn{1}{r|}{(\textbf{0.00})(\textbf{0.09})}   & \multicolumn{1}{r|}{(0.00)(0.00)}                    & \multicolumn{1}{r|}{(\textbf{0.00})(\textbf{0.09})}   & \multicolumn{1}{r|}{(0.00)(0.00)}                    & \multicolumn{1}{r|}{(\textbf{0.00})(\textbf{0.09})}   & \multicolumn{1}{r|}{(0.00)(0.00)}                    & (\textbf{0.00})(\textbf{0.09  })  \\
			mem\_ctrl                   & \multicolumn{1}{r|}{145983.75} & \multicolumn{1}{r|}{15931.2}    & \multicolumn{1}{r|}{157320.88}  & \multicolumn{1}{r|}{15218.73}  & \multicolumn{1}{r|}{(-5.00)(\textbf{2.57})}          & \multicolumn{1}{r|}{(-3.39)(-8.23)}                   & \multicolumn{1}{r|}{(-4.01)(\textbf{3.48})}          & \multicolumn{1}{r|}{(\textbf{3.49})(-1.02)}           & \multicolumn{1}{r|}{(-3.91)(\textbf{3.58})}          & \multicolumn{1}{r|}{(\textbf{1.85})(-2.74)}           & \multicolumn{1}{r|}{(-3.82)(\textbf{3.66})}          & (-3.39)(-8.23 )                   \\
			router                      & \multicolumn{1}{r|}{1017.23}   & \multicolumn{1}{r|}{3139.16}    & \multicolumn{1}{r|}{1151.1}     & \multicolumn{1}{r|}{2167.14}   & \multicolumn{1}{r|}{(-0.61)(\textbf{11.09})}         & \multicolumn{1}{r|}{(\textbf{20.15})(-15.66)}         & \multicolumn{1}{r|}{(\textbf{0.12})(\textbf{11.74})} & \multicolumn{1}{r|}{(\textbf{20.37})(-15.34)}         & \multicolumn{1}{r|}{(-1.60)(\textbf{10.22})}         & \multicolumn{1}{r|}{(\textbf{19.41})(-16.74)}         & \multicolumn{1}{r|}{(\textbf{0.37})(\textbf{11.96})} & (\textbf{20.15})(-15.66)          \\ \hline
			log2                        & \multicolumn{1}{r|}{106245.65} & \multicolumn{1}{r|}{37328.95}   & \multicolumn{1}{r|}{128648.38}  & \multicolumn{1}{r|}{35983.87}  & \multicolumn{1}{r|}{(-0.85)(\textbf{16.71})}         & \multicolumn{1}{r|}{(\textbf{3.91})(\textbf{0.32})}   & \multicolumn{1}{r|}{(-1.84)(\textbf{15.90})}         & \multicolumn{1}{r|}{(\textbf{4.27})(\textbf{0.69})}   & \multicolumn{1}{r|}{(-2.03)(\textbf{15.74})}         & \multicolumn{1}{r|}{(\textbf{2.36})(-1.29)}           & \multicolumn{1}{r|}{(-2.38)(\textbf{15.45})}         & (\textbf{3.91})(\textbf{0.32})    \\
			square                      & \multicolumn{1}{r|}{67833.8}   & \multicolumn{1}{r|}{17163.04}   & \multicolumn{1}{r|}{69424.08}   & \multicolumn{1}{r|}{13962.72}  & \multicolumn{1}{r|}{(\textbf{4.68})(\textbf{6.86})}           & \multicolumn{1}{r|}{(-6.77)(-31.24)}                  & \multicolumn{1}{r|}{(\textbf{4.75})(\textbf{6.94})}  & \multicolumn{1}{r|}{(-4.87)(-28.91)}                  & \multicolumn{1}{r|}{(\textbf{4.68})(\textbf{6.86})}  & \multicolumn{1}{r|}{(-5.46)(-29.63)}                  & \multicolumn{1}{r|}{(\textbf{4.49})(\textbf{6.68})}  & (-6.77)(-31.24)                   \\
			adder                       & \multicolumn{1}{r|}{4082.67}   & \multicolumn{1}{r|}{16195.36}   & \multicolumn{1}{r|}{4819.62}    & \multicolumn{1}{r|}{12349.82}  & \multicolumn{1}{r|}{(-22.59)(-3.84)}                 & \multicolumn{1}{r|}{(\textbf{-15}.00)(-50.81)}        & \multicolumn{1}{r|}{(-22.59)(-3.84)}                 & \multicolumn{1}{r|}{(-15.00)(-50.81)}                 & \multicolumn{1}{r|}{(-22.59)(-3.84)}                 & \multicolumn{1}{r|}{(-14.99)(-50.80)}                 & \multicolumn{1}{r|}{(-21.97)(-3.32)}                 & (-15.00)(-50.81)                  \\
			sin                         & \multicolumn{1}{r|}{21080.22}  & \multicolumn{1}{r|}{18431.83}   & \multicolumn{1}{r|}{26298.97}   & \multicolumn{1}{r|}{20669.35}  & \multicolumn{1}{r|}{(-8.31)(\textbf{13.18})}         & \multicolumn{1}{r|}{(\textbf{9.09})(\textbf{18.93})}  & \multicolumn{1}{r|}{(-2.30)(\textbf{18.00})}         & \multicolumn{1}{r|}{(\textbf{5.43})(\textbf{15.67})}  & \multicolumn{1}{r|}{(-4.82)(\textbf{15.98})}         & \multicolumn{1}{r|}{(-0.81)(\textbf{10.11})}          & \multicolumn{1}{r|}{(-0.93)(\textbf{19.10})}         & (\textbf{9.09})(\textbf{18.93})   \\
			div                         & \multicolumn{1}{r|}{237853.11} & \multicolumn{1}{r|}{412158.44}  & \multicolumn{1}{r|}{309789.59}  & \multicolumn{1}{r|}{363934.53} & \multicolumn{1}{r|}{(-10.87)(\textbf{14.87})}        & \multicolumn{1}{r|}{(\textbf{1.01})(-12.11)}          & \multicolumn{1}{r|}{(-4.54)(\textbf{19.73})}         & \multicolumn{1}{r|}{(\textbf{9.59})(-2.39)}           & \multicolumn{1}{r|}{(-5.57)(\textbf{18.94})}         & \multicolumn{1}{r|}{(\textbf{9.85})(-2.09)}           & \multicolumn{1}{r|}{(-4.77)(\textbf{19.56})}         & (\textbf{1.01})(-12.11)           \\
			hyp                         & \multicolumn{1}{r|}{820478.12} & \multicolumn{1}{r|}{2887324.25} & \multicolumn{1}{r|}{970732.25}  & \multicolumn{1}{r|}{2507905.5} & \multicolumn{1}{r|}{(-7.60)(\textbf{9.05})}          & \multicolumn{1}{r|}{(\textbf{22.22})(\textbf{10.45})} & \multicolumn{1}{r|}{(-7.74)(\textbf{8.94})}          & \multicolumn{1}{r|}{(\textbf{23.28})(\textbf{11.68})} & \multicolumn{1}{r|}{(-7.82)(\textbf{8.87})}          & \multicolumn{1}{r|}{(\textbf{23.81})(\textbf{12.29})} & \multicolumn{1}{r|}{(-8.46)(\textbf{8.32})}          & (\textbf{22.22})(\textbf{10.45})  \\
			max                         & \multicolumn{1}{r|}{11596.12}  & \multicolumn{1}{r|}{25593.73}   & \multicolumn{1}{r|}{12559.54}   & \multicolumn{1}{r|}{17387.36}  & \multicolumn{1}{r|}{(-3.54)(\textbf{4.40})}          & \multicolumn{1}{r|}{(\textbf{24.41})(-11.26)}         & \multicolumn{1}{r|}{(-2.10)(\textbf{5.73})}          & \multicolumn{1}{r|}{(\textbf{26.93})(-7.56)}          & \multicolumn{1}{r|}{(-1.92)(\textbf{5.90})}          & \multicolumn{1}{r|}{(\textbf{27.57})(-6.62)}          & \multicolumn{1}{r|}{(-2.03)(\textbf{5.80})}          & (\textbf{24.41})(-11.26)          \\
			sqrt                        & \multicolumn{1}{r|}{108308.88} & \multicolumn{1}{r|}{1034090.5}  & \multicolumn{1}{r|}{198626.75}  & \multicolumn{1}{r|}{872972.5}  & \multicolumn{1}{r|}{(-44.80)(\textbf{21.04})}        & \multicolumn{1}{r|}{(\textbf{4.92})(-12.62)}          & \multicolumn{1}{r|}{(-37.14)(\textbf{25.22})}        & \multicolumn{1}{r|}{(\textbf{19.14})(\textbf{4.22})}  & \multicolumn{1}{r|}{(-45.82)(\textbf{20.49})}        & \multicolumn{1}{r|}{(\textbf{17.12})(\textbf{1.83})}  & \multicolumn{1}{r|}{(-32.15)(\textbf{27.94})}        & (\textbf{4.92})(-12.62)           \\
			multiplier                  & \multicolumn{1}{r|}{99859.52}  & \multicolumn{1}{r|}{24972.81}   & \multicolumn{1}{r|}{119304.42}  & \multicolumn{1}{r|}{19641.37}  & \multicolumn{1}{r|}{(-3.10)(\textbf{13.71})}         & \multicolumn{1}{r|}{(\textbf{13.34})(-10.18)}         & \multicolumn{1}{r|}{(\textbf{2.76})(\textbf{18.61})} & \multicolumn{1}{r|}{(\textbf{9.72})(-14.78)}          & \multicolumn{1}{r|}{(\textbf{3.38})(\textbf{19.12})} & \multicolumn{1}{r|}{(\textbf{14.09})(-9.23)}          & \multicolumn{1}{r|}{(\textbf{4.51})(\textbf{20.08})} & (\textbf{13.34})(-10.18)          \\
			bar                         & \multicolumn{1}{r|}{10896.7}   & \multicolumn{1}{r|}{4797.36}    & \multicolumn{1}{r|}{10896.7}    & \multicolumn{1}{r|}{4816.15}   & \multicolumn{1}{r|}{(0.00)(0.00)}                    & \multicolumn{1}{r|}{(-0.39)(0.00)}                    & \multicolumn{1}{r|}{(0.00)(0.00)}                    & \multicolumn{1}{r|}{(-0.39)(0.00)}                    & \multicolumn{1}{r|}{(0.00)(0.00)}                    & \multicolumn{1}{r|}{(-0.39)(0.00)}                    & \multicolumn{1}{r|}{(0.00)(0.00)}                    & (-0.39)(0.00  )                   \\ \hline
			\textbf{Average}            & \multicolumn{1}{r|}{/}        & \multicolumn{1}{r|}{/} & \multicolumn{1}{r|}{(-16.37)(0.0)}     & \multicolumn{1}{r|}{(8.82)(0.0)}                    & \multicolumn{1}{r|}{(-10.72)(\textbf{4.13})}         & \multicolumn{1}{r|}{(\textbf{10.10})(-0.69)}          & \multicolumn{1}{r|}{(-9.11)(\textbf{5.38})}          & \multicolumn{1}{r|}{(\textbf{8.44})(-2.20)}           & \multicolumn{1}{r|}{(-9.89)(\textbf{4.85})}          & \multicolumn{1}{r|}{(\textbf{8.64})(-1.98)}           & \multicolumn{1}{r|}{(-9.00)(\textbf{5.39})}          & (\textbf{10.10})(-0.69)           \\ \hline
		\end{tabular}
	}
\end{table*}

\begin{table}[t]
	\centering
	\caption{Our method compares with ABC using supergate, but not limiting the number of supergate attempts of cut.}
	\renewcommand\tabcolsep{19.4pt}
	\label{table:ultra}
	\begin{tabular}{|c|c|c|}
		\hline
		Benchmarks & Area imp\% & Delay imp\% \\ \hline
		1 thread   & 4.65       & 5.32        \\ \hline
		2 threads  & 4.63       & 5.28        \\ \hline
		4 threads  & 4.62       & 5.17        \\ \hline
		8 threads  & 4.60       & 5.29        \\ \hline
		16 threads & 4.67       & 5.16        \\ \hline
		32 threads & 4.57       & 5.30        \\ \hline
	\end{tabular}
\end{table}

\subsection{Setup}
\textbf{Environment.} 
We implemented the proposed method in C++ language under the environment of the GCC 11.1.0 x86, and the 64-bit Ubuntu 18.04.6 operation system. 
We collected our experimental results on a system with 128G shared RAM and 2 Intel® Xeon® Gold 6252 CPU @ 2.10GHz processors, each with 48 physical cores.

\textbf{Benchmarks.}
We perform experiments on EPFL benchmark\cite{benchmark}. 
This benchmark contains 10 arithmetic circuits and 10 control/random circuits with circuit sizes ranging from 0.1k to 210k.
The standard cell library is using the opensource sky130~(sky130\_fd\_sc\_hd\_\_tt\_025C\_1v80)~\cite{sky130}, which contains 82 useful gates for supergates computing.

\textbf{Parameters.}
For the parameters of supergates computing, the maximum input number of the library is 5, and the level of supergates is 2.
For ASIC technology mapping, the cut enumeration limit is 24, and each cut tries 30 supergates.

\subsection{Evaluation on Runtime Reduction}

\begin{figure}[t]
    \centering
    \includegraphics[width=0.95\linewidth]{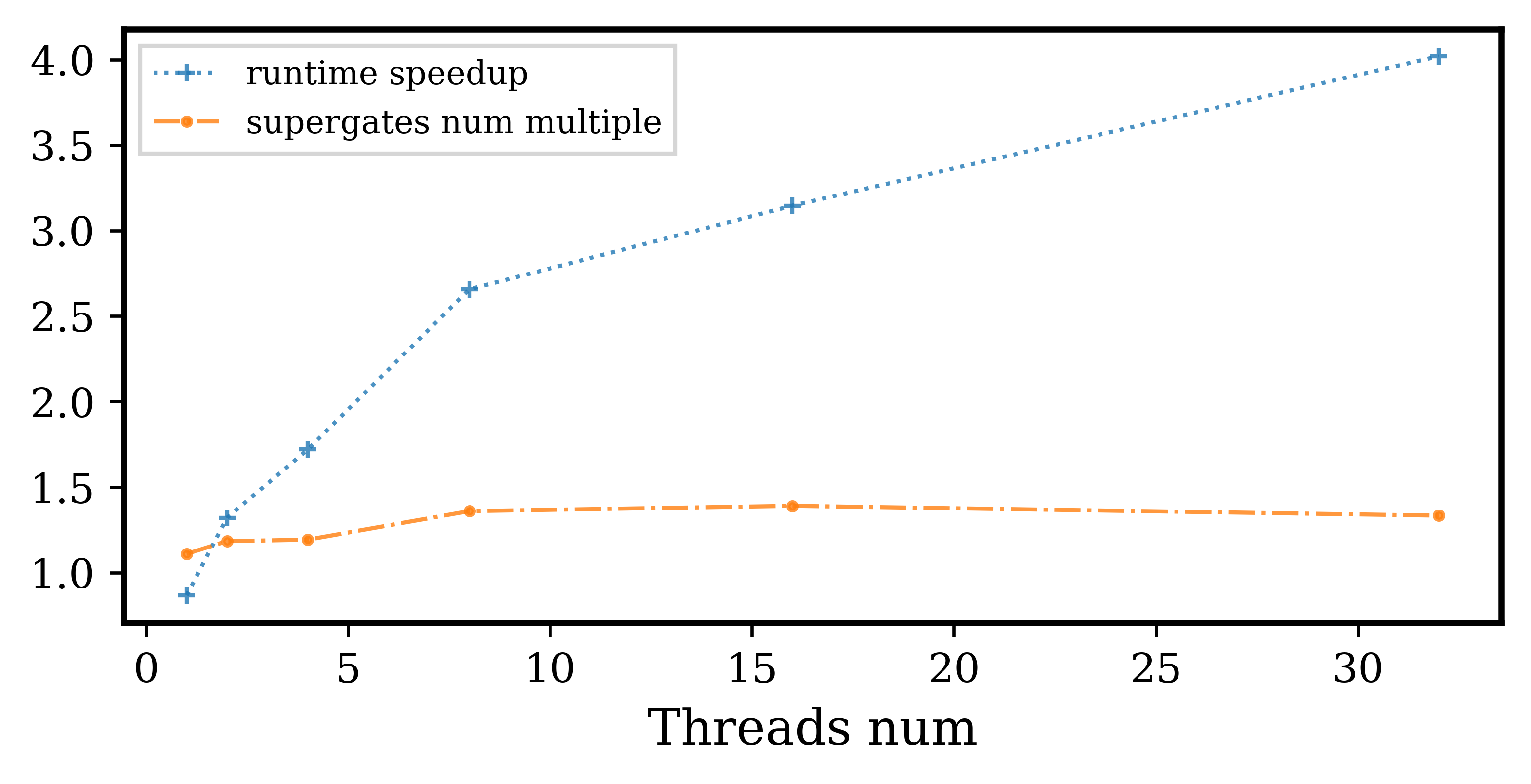}
    \caption{Our method compared with ABC, the runtime speedup and the multiple of number of supergates.}
    \label{fig:experiment-speedup}
\end{figure}

To demonstrate the scalability of parallel supergates computing, we present in \figref{fig:experiment-speedup} the comparison of the number and time of supergates generated by our method with ABC under different threads, the speedup is our method run time divided by the runtime of ABC supergates computing, which is 100s.
The above results were performed 20 times under each thread, and the average was taken for display to eliminate external noise effects.

From \figref{fig:experiment-speedup}, it can be seen that the supergate computing time of ABC was fixed at 100 seconds, resulting in 6050082 supergates being generated. 
And our method can generate 6710591 supergates in 115.38 seconds when the number of threads is 1, with speed similar to ABC.
However, our method can generate more supergates when the number of threads is higher, and the number of supergates generated may vary in a multi-threaded environment.

\subsection{Evaluation on QoR Improvement}

To demonstrate the optimization effect of our method on the subsequent ASIC technology mapping results, the \cref{table: QoR} shows the optimization percentages of the area and delay for our method under different threads compared to the results of ABC without and with supergate. 
A positive number indicates a higher optimization effect than ABC, while a negative number indicates the opposite.
For the output netlist, we use \textit{stime} command of ABC to calculate its area and delay.
Compared to the ``ABC without sg'', the proposed method can attain mostly 10.1\% in delay reduction and 9\% in area increase; while the ``ABC with sg'' is with 8.82\% in delay reduction and 16.37\% area increase.
Compared to ``ABC with sg'', the proposed method reduces area by 5.39\% with only a 0.69\% increase in delay.

For the purpose of saving mapping time, the number of supergates attempted by each cut was limited, but this also limited the space for supergates to play. 
Therefore, The \cref{table:ultra} lists the optimization effect that the mapping result can achieve without restriction. This result is compared with the result of ABC using supergate.
From the above two tables, we can see that our method has great potential for optimizing the delay of mapping results.


\section{Conclusion}
\label{sec:conclusion}


In this paper, we focus on parallel supergates computing, which can enhance ASIC technology mapping.
We first find the conflict between the computation cost and delay reduction of the supergates generation.
Then we proposed a parallel supergate computing algorithm that consists of three steps: pre-processing, input-constrained parallel candidates computing, and post-processing.
The combination of recursion and backtracking is used to enumerate candidates, which effectively avoids the storage of intermediate variables and saves space.
Finally, the sufficient experiments show the efficiency of our proposed method.

\section*{Acknowledgment}
This work is supported by the Major Key Project of PCL (No. PCL2023A03).

\bibliographystyle{IEEEtran}

\bibliography{supergates.bib}  

\begin{thebibliography}{10}
\providecommand{\url}[1]{#1}
\csname url@samestyle\endcsname
\providecommand{\newblock}{\relax}
\providecommand{\bibinfo}[2]{#2}
\providecommand{\BIBentrySTDinterwordspacing}{\spaceskip=0pt\relax}
\providecommand{\BIBentryALTinterwordstretchfactor}{4}
\providecommand{\BIBentryALTinterwordspacing}{\spaceskip=\fontdimen2\font plus
\BIBentryALTinterwordstretchfactor\fontdimen3\font minus \fontdimen4\font\relax}
\providecommand{\BIBforeignlanguage}[2]{{%
\expandafter\ifx\csname l@#1\endcsname\relax
\typeout{** WARNING: IEEEtran.bst: No hyphenation pattern has been}%
\typeout{** loaded for the language `#1'. Using the pattern for}%
\typeout{** the default language instead.}%
\else
\language=\csname l@#1\endcsname
\fi
#2}}
\providecommand{\BIBdecl}{\relax}
\BIBdecl

\bibitem{para2}
M.~Shen and G.~Luo, ``Accelerate {FPGA} routing with parallel recursive partitioning,'' in \emph{2015 IEEE/ACM International Conference on Computer-Aided Design (ICCAD)}, 2015, pp. 118--125.

\bibitem{para3}
L.~Stok, ``Developing parallel {EDA} tools {[The Last Byte]},'' \emph{IEEE Design \& Test}, vol.~30, no.~1, pp. 65--66, 2013.

\bibitem{para4}
V.~Possani, Y.-S. Lu, A.~Mishchenko, K.~Pingali, R.~Ribas, and A.~Reis, ``Unlocking fine-grain parallelism for {AIG} rewriting,'' in \emph{2018 IEEE/ACM International Conference on Computer-Aided Design (ICCAD)}, 2018, pp. 1--8.

\bibitem{lasted-GPU-unlock}
L.~Li, R.~Li, and Y.~Ha, ``A recursion and lock free {GPU}-based logic rewriting framework exploiting both intra-node and inter-node parallelism,'' \emph{IEEE Transactions on Computer-Aided Design of Integrated Circuits and Systems}, pp. 1--1, 2023.

\bibitem{liu2023rethinking}
T.~Liu and E.~F.~Y. Young, ``Rethinking {AIG} resynthesis in parallel,'' in \emph{60th ACM/IEEE Design Automation Conference (DAC)}, 2023.

\bibitem{chatterjee2007algorithms}
S.~Chatterjee, \emph{On algorithms for technology mapping}.\hskip 1em plus 0.5em minus 0.4em\relax University of California, Berkeley, 2007.

\bibitem{para1}
S.~R. Whiteley and J.~Kawa, ``Progress toward {VLSI}-capable {EDA} tools for superconductive digital electronics,'' in \emph{2019 IEEE International Superconductive Electronics Conference (ISEC)}, 2019, pp. 1--3.

\bibitem{ASIC-time}
F.~Abid and N.~Izeboudjen, ``Technology-independent approach for {FPGA} and {ASIC} implementations,'' in \emph{2015 4th International Conference on Electrical Engineering (ICEE)}, 2015, pp. 1--4.

\bibitem{ASIC-time2}
X.~Wang, M.~Yang, Z.~Li, and L.~Wang, ``Parallelized technology mapping to general {PLB}s by adaptive circuit partitioning,'' in \emph{2021 International Conference on Field-Programmable Technology (ICFPT)}, 2021, pp. 1--5.

\bibitem{tree-mapping}
K.~Keutzer, ``{DAGON}: Technology binding and local optimization by {DAG} matching,'' in \emph{24th ACM/IEEE Design Automation Conference (DAC)}, 1987, pp. 341--347.

\bibitem{priority-cut}
A.~Mishchenko, S.~Cho, S.~Chatterjee, and R.~Brayton, ``Combinational and sequential mapping with priority cuts,'' in \emph{2007 IEEE/ACM International Conference on Computer-Aided Design (ICCAD)}, 2007, pp. 354--361.

\bibitem{supergate-2005}
A.~Mishchenko, S.~Chatterjee, R.~Brayton, X.~Wang, and T.~Kam, ``Technology mapping with boolean matching, supergates and choices,'' \emph{Berkeley Logic Synthesis and Verification Group}, 2005.

\bibitem{benchmark}
L.~Amarú, P.-E. Gaillardon, and G.~D. Micheli, ``{The EPFL Combinational Benchmark Suite},'' in \emph{Proceedings of the 24th International Workshop on Logic {\&} Synthesis (IWLS)}, 2015.

\bibitem{sky130}
{Google}, ``{SkyWater-PDK},'' \url{https://github.com/google/skywater-pdk}, 2021, accessed: February 1, 2024.

\end{thebibliography}

\end{CJK*}
\end{document}